# Time-Assisted Authentication Protocol

[1]Muhammad Bilal

University of Science and Technology, Korea

Electronics and Telecommunication Research Institute, Rep. of Korea

mbilal@etri.re.kr, engr.mbilal@yahoo.com

[2] Shin-Gak Kang

Electronics and Telecommunication Research Institute, Rep. of Korea

sgkang@etri.re.kr,

**Abstract**
Authentication is the first step toward establishing a service provider and customer (C-P) association. In a mobile network environment, a lightweight and secure authentication protocol is one of the most significant factors to enhance the degree of service persistence. This work presents a secure and lightweight keying and authentication protocol suite termed TAP (Time-Assisted Authentication Protocol). TAP improves the security of protocols with the assistance of time-based encryption keys and scales down the authentication complexity by issuing a re-authentication ticket. While moving across the network, a mobile customer node sends a re-authentication ticket to establish new sessions with service-providing nodes. Consequently, this reduces the communication and computational complexity of the authentication process. In the keying protocol suite, a key distributor controls the key generation arguments and time factors, while other participants independently generate a keychain based on key generation arguments. We undertake a rigorous security analysis and prove the security strength of TAP using CSP and rank function analysis.

## I. Introduction

Newly emerging technologies such as IoT devices, smartphones with various sensors, wireless power charging systems, wearable devices, and smart sensors have brought forth different services with new types of service provider and customer (P-C) association requirements. For any service provider and customer (P-C) association, authentication is the first step. Typically, customers seek to obtain services from an authenticated service provider, and service providers are concerned about providing services to verified customers. However, to provide seamless service in a mobile environment, the service provider entity should be able to authenticate the mobile customer node with only a minimum delay time. This implies that the message exchange complexity of the authentication process should be low. A suitable authentication protocol for such an application should be lightweight and yet should ensure a proper protocol security level. Moreover, a secure authentication protocol certifies that the communicating entity is an authorized entity which is alive and participating in a protocol run according to a defined role. Further, the protocol run follows the correct pre-set sequence of a protocol run, and this should be achieved over insecure communication channels between the service provider and customer nodes [1-2].

Various authentication protocols have been designed to meet the requirements of different applications, including authentication for sensor networks [3], authentication for streaming data [4], authentication for IoT solutions [5], and authentication for ad-hoc networks [6]. The protocols developed thus far can be categories into the three primary categories of password-based authentication [7-8], certificate-based authentication [9-10], and signature-based authentication protocols [11]. In this study, we introduce a novel, lightweight, and secure time-assisted authentication protocol suite termed TAP [12]. The TAP protocol suite consists of keying and authentication protocol suites. The keying protocol suite consists of key agreement and key retrieval protocols. Likewise, the authentication protocol consists of the three protocols of the initial authentication protocol, re-authentication protocol 1, and re-authentication protocol 2. The initial authentication protocol is a password-based authentication protocol, whereas the re-authentication protocols are certificate-based authentication protocols where the certificate is a time-based ticket issued for a specified time duration.

The well-known authentication protocol Kerberos [13] also employs a ticket for authentication, though in contrast with Kerberos, TAP does not require time synchronization. Moreover, in TAP the ticket verifier can authenticate the customer by itself without contacting the ticket-granting entity. Moreover, a stolen ticket in TAP is of no significance, while in the case of Kerberos, an unexpired stolen ticket is useful for an intruder. In an earlier study [14], the author analyzed the Kerberos protocol using CSP (communicating sequential processes) and a rank function analysis and established that the protocol is vulnerable to few known attacks, whereas our CSP and rank function analysis show that TAP is an entirely secure protocol. To ensure secure authentication, TAP employs a distributed keychain generation mechanism and reverses the time-based usage of the keychain. Unlike TESLA [15], in TAP the keychain is independently generated by multiple devices and is also used by each device to drive other encryption keys. For instance, during an interval $i$ when a customer node wants to obtain service, it acquires an authentication ticket encrypted with the $ith$ key, and once it moves across the network, the verifier can verify the ticket by decrypting the ticket using the $ith$ key. The key distribution and time-based usage of the keychain are controlled by the main authentication entity (ME), which can be an independent authentication server or an agent installed on a server. While moving across the network, a mobile customer experiences the authentication process multiple times. With authentication ticket, the overall computational and message exchange complexity of the authentication process is reduced significantly.

To establish the overall infrastructure for service and authentication, each main authentication entity (ME) creates a group of service provider entities and neighboring main authentication entities (ME). Using group key encryption, the main authentication entity (ME) broadcasts the time-based key generation parameters to all group members. In earlier work [16], the author presented and examined several group key management protocols. In this discussion, we assume that the main authentication entity (ME) has knowledge of the public keys of all group members; hence, it can share a group key with new group members using public-private key semantics. To determine the security strength of TAP, we considered an intruder as discussed in earlier work [17] with certain extra capabilities. For instance, the intruder is capable of operating on all communication channels between the customer and service provider entity. Moreover, the intruder can redirect, spoof, replay or block messages, and has initially known information; e.g., it knows the IDs of all users and service provider entities. It can intercept, record, and generate a message from known information. Given the presence of such an intruder, we discuss the strength of TAP against some known attacks and instances of intercepted credentials, such as impersonation of a customer node, impersonation of a service provider entity, and a replay attack. We further analyze TAP using CSP and rank function analysis.

The remainder of the paper is organized as follows: Section II gives a brief system overview and discusses the proposed strategy in particular. In Section III, we review the strength of TAP against several known attacks and cases of intercepted credentials. Next, the formal analysis of TAP using CSP and rank function analysis is presented in Section IV. Then, in Section V, we compare the security and performance of the TAP protocol suite with those of previous methods work in the literature. Finally, we provide concluding remarks in Section VI.

## II. Analytical Model and Proposed Scheme

The TAP system is not limited to a particular network type or application; it is suitable for sensor networks, mobile networks, and client-server applications, among others. For the discussion and analysis, we consider that the TAP system consists of the three major entities of the main authentication entity ($ME$), the service provider entity ($P$), and the customer node ($C$). **The main authentication entity ($ME$)** knows the public keys of all service provider entities ($P$) and customer nodes ($C$). It authenticates and issues a time-based ticket ($T_k$) for re-authentication. It also authenticates any newly joining service provider entity ($P$). During the authentication process of $P$, the $ME$ sends the customer node's requirement profile for an efficient and optimized $P \leftrightarrow C$ relationship. In addition, the $ME$ controls the key distribution and derives various keys from the keychain. All $MEs$ are connected via secure links and share public keys with all of their neighbors. **The service provider entity ($P$)** authenticates $C$ upon receiving a valid authentication ticket ($T_k$) and starts providing services based upon the user profile. Upon the receipt of the initial authentication request, $P$ forwards the message to $ME$. In some applications, $P$ has two separate areas of operation: the service-delivery area ($A_s$) and the communication area ($A_c$). $P$ may authenticate or forward a request to $ME$ while $C$ is in $A_s$, and it may provide services once it enters $A_c$. **The customer node ($C$)** can join or leave the system dynamically and can move across the network. Either $C$ joins the network or switches from one service provider ($P$) to another. In both cases, $C$ is responsible for initiating the authentication procedure.

**Notations:**
- $ME_j$ = The $jth$ main authentication entity.
- $P_j$ = The $jth$ service provider entity.
- $C_i$ = The $ith$ customer node.

- $A\Delta B$ = A is associated with B such that B is in controlling authority.
  - $P_i \Delta ME_i$ represents that $ith$ service provider entity is associated with $jth$ ME.
    - $G_j$ =Group of all associated entities of $jth$ ME.
    - $G_j^o$ = Group of all non-associated entities of $jth$ ME who knows the $K_G^j$.
- $K_0^j$ =The time-based key generated at $0th$ interval known as commitment key.
- $K_i^j$ =The time-based key generated at $ith$ interval by $ME_j$.
- $K_C^j$ = The secret public key of $jth$ Customer, it is publically known to the $ME$s only.
- $K_{ME}^j$ is the public key of $ME_j$, $K_P^j$ is the public key of $P_j$ and $K_C^j$ is the public key of $C_j$.
- $K_G^j$ =A group key generated by $ME_j$.
- $K_S^i$ = A C→P session key generated at $ith$ interval.
- $K_k^p$ = A partial key of $C_k$ used to generate $C \rightarrow C$ session key.
- $K_{i,j}$ = A C→C session key generated from partial keys.
- $E_B^i(m)$ =Encrypting message 'm' with key $K_B^i$.
- $E_{ME}^j(m)$ represents the encryption of 'm' using $K_{ME}^j$.
- $V_0$ =The index value for interval 0.
- $T_k$ = The $kth$ Ticket generated at $ith$ interval.
- $Z(A)$ = An intruder $Z$ impersonating entity $A$.
- $N_i$ = $ith$ nonce in a TAP message exchange.

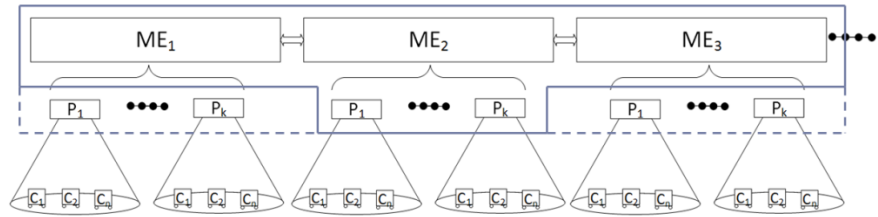

Figure 1. System architecture

The overall system architecture is shown in Figure 1. Each $ME$ creates a group which consists of neighbor $ME$s and associated service provider entities. The $ME$ shares a symmetric group key $K_G^j$ with all group members. For instance, in Figure 1 $ME_2$ creates a group consisting of $ME_1$, $ME_3$, and associated service providers $(P_1 \dots P_k)$. The group formation and group key $K_G^j$ management processes can be done as per system constraints and requirements. In this discussion, we assume that $ME_j$ knows the public keys of all group members; consequently, it can share group key $K_G^j$ with new group members using public-private key encryption. The group key is a rational choice given the assumption that group members are not synchronized with regard to time. In a stable synchronized distributed system, the group key $K_G^j$ can be replaced by the time-based key $K_i^j$, where $K_i^j$ should be the final key in the keychain. The TAP scheme consists of two protocol suites, the keying protocol suite and the authentication protocol suite, which are discussed below in subsections A and B, respectively

### B. Keying Protocol:

As in TESLA [15], TAP also generates a keychain using an irreversible function. However, in TESLA keys are used to authenticate a broadcasting entity; in TAP the keys are used to derive several other keys, such as a ticket encryption key, a C↔P session key, and partial session key generators which provide security to the authentication system overall. Moreover, in TESLA [15], the keychain is generated in the broadcasting entity, whereas in TAP all members of $G_j$ generate and share the same "chain of key generators" of length $L$. The group leader ($ME_j$) shares the key generation information ($Key_{MSG}$) with group members when they join the group. After the expiration of $T_d$ (at a valid time for commitment key) $ME_j$ broadcast $Key_{MSG}$ to all members of $G_j$; $Key_{MSG}$ is encrypted with $K_G^j$ or in the case of a time-synchronized system is encrypted with $K_l^j$ (the final key in the keychain). The key generation information ($Key_{MSG}$) is used to generate a commitment key ($K_0^j$); afterward, all group members independently generate the "chain of key generators." The group leader $ME_j$ controls the procedure of commitment key generation and all related characteristics.

The key generation information $Key_{MSG} = lookup(I, O) \,||\, T_d \,||\, T_c^j \,||\, N_0^j \,||\, L \,||\, MODE$ consists of several pieces of information. $I$ and $O$ are index and offset values, respectively, which are used to select a predefined value from the secret $TABLE$. $T_d$ is the valid time duration for commitment key $K_0^j$, $T_C$ is the time on $ME_j$ when it was broadcasting $Key_{MSG}$, $N_o^j$ is a nonce and $MODE$ is the key retrieval mode (discussed in subsection b). Note that $T_d$ is divided into $L$ number of intervals; it also determines the keychain length.

#### a) Key Generation and Distribution:

The time frame shown in Figure 3-(a) is composed of three periods: the time required for key generation ($T_G$), the time needed for key distribution ($T_{Dis}$), and the valid time for the commitment key ($T_d$). All $G_i$ members follow the subsequent steps: (1) run a function $g$ (shown in Figure 3-(b)) to generate the commitment key generator $G_0^j = g(lookup(I, O), T_d, N_0^j)$. (2) An irreversible function $F$ takes $G_0^j$ as input argument and generates a "chain of key generators" of length $L$, $\{F(G_0^j) = G_1^j, F(G_1^j) = G_2^j \ldots \ldots F(G_{l-1}^j) = G_l^j$; i-e $F(G_k^j)^i = G_{k+i}^j$. (3) Using another irreversible function $f$ (shown in Figure 3(c)) all members of $G_j$ generate an index value and keychain, $f(G_0^j) = (K_0^j||V_0), f(G_1^j) = (K_1^j||V_1) \ldots f(G_l^j) = (K_l^j||V_l) \rightarrow V||K$; where, $V = \{V_0, V_1, \ldots V_l\}$ and $K = \{K_0^j, K_1^j, \ldots K_l^j\}$. These keys are used for ticket encryption; for instance, the $ith$ key ($K_i^j$) is used to encrypt tickets ($T_k$) issued during the $ith$ time interval. Keys are disclosed in reverse order such the intruder cannot generate future keys, whereas the indexing vector ($V$) serves in an indexing role to retrieve the ticket encryption key. For instance, if $V_i$ corresponds to the $ith$ interval, the key is retrieved as $F(G_0^j)^i = G_i^j \Rightarrow f(G_i^j) = (K_i^j, V_i)$. (4) $ME_j$ broadcasts $Key_{MSG}$ for the next chain to all group members using $K_G^j$ or $K_i^i$. Steps 1-3 are performed in $T_G$ and step 4 is executed during $T_{Dis}$. In addition to the time-based keys (for ticket encryption), $ME_j$ also generates a unique C↔P session key $K_S^i = H(K_i^j||H(K_C^k))$ for verified $C_k$ and $P_j$. Moreover, the validity time for $K_S^k$ is given as $T_R = l_i - L$.

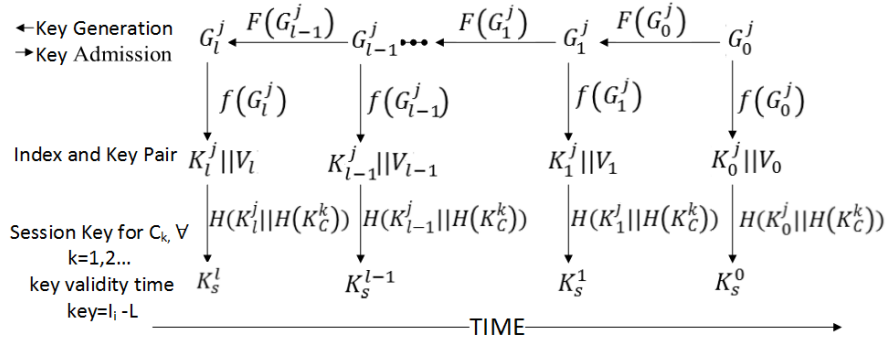

Figure 2. Time-based keys generation and admission with reference to time passage.

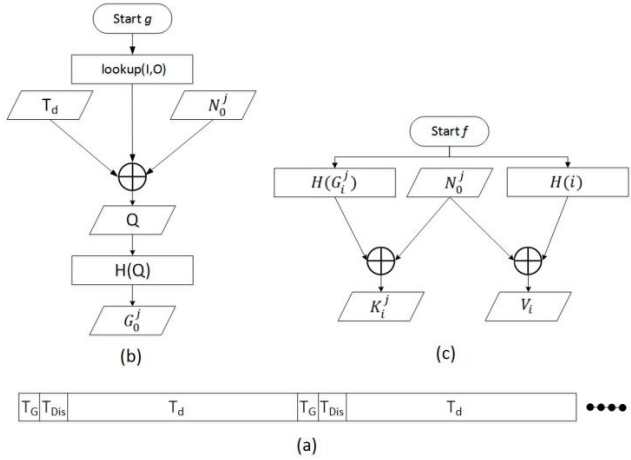

Figure 3. (a) Time frame. (b) Structure of function $g \Rightarrow H(lookup(I,O) \oplus T_d \oplus N_0^j)$. (c) Structure of function $f \Rightarrow H(G_i^j) \oplus N_0^j ||H(i) \oplus N_0^j)$.

b) The Retrieval Modes For The Ticket Encryption Key:

The authentication ticket consists of two segments: the customer information segment and the key retrieval information segment respectively encrypted with the time-based key $K_l^i$ and group key $K_G^j$. The key retrieval procedure depends on the structure of the ticket, which is determined by the system requirements and constraints. We propose three different key retrieval modes, as presented below.

$E_i^j(C_k||K_s^i||V_i||Profile||H_{head})||E_G^j(C_k||V_i||H(G_i)||H(K_C^k))$ ∴ Mode-1

$E_i^j(C_k||K_s^i||V_i||Profile||H_{head})||E_G^j(C_k||(T_t)||H(G_i)||H(K_C^k))$ ∴ Mode-2

$E_i^j(C_k||K_s^i||V_i||Profile||H_{head})||E_G^j(C_k||V_i||H_{head}||Vector_{Hash}||H(K_C^k))$ ∴ Mode-3

The second half of $T_k$ depends on the mode and consists of time-based key retrieval information; once the time-based key is retrieved, it is then employed to decrypt and verify the first half of $T_k$. In **mode-01**, $ME_j$ adds the index value $(V_i)$ in the key retrieval information segment. The ticket verifier compares the appended value within the locally generated vector $(V)$. A match at the $ith$ place indicates that the ticket is generated by $ME_j$ at the $ith$ interval and can be decrypted by key $K_i^j$. In **mode-02**, $ME_j$ inserts the ticket issuing time $T_t$ into the key retrieval information segment. The verifiers can search for the value of the index within the following range:

$[|T_c^i - T_t| * \frac{T_d}{L} - \epsilon_{updated}, |T_c^i - T_t| * \frac{T_d}{L} + \epsilon_{updated}]$

Where, $T_c^i$ is the current time of the verifier clock and $\epsilon$ is the time drift. Initially, the value of $\epsilon$ is calculated as $\epsilon_0 = |T_c^i - T_c^j|$.

Upon each successful retrieval of $K_k^j$, the value of $\epsilon$ is updated as shown below.

$\epsilon_{updated} = w_0 * \epsilon_{previous} + w_1 * \epsilon_{current}$

In this equation, $w_0 = 1 - \frac{T_c^i + T_t}{T_d}$ and $w_1 = \frac{T_c^i + T_t}{T_d}$

In **mode-03**, all $G_i$ members independently generate a binary hash tree whose leaf nodes are 'indexing values' taken from index vector $(V)$. $ME_j$ adds the index value $(V_i)$ and $\log_2|V|$ number of hash values in key retrieval information segment; these hash values are selected nodes of the hash tree. Likewise, the verifier can reconstruct the hash tree with the total $\log_2|V|$ number of hash operations, which gives the complexity of $O(log)$. After the reconstruction of the tree and confirmation of the head node, the verifier retrieves the index value by running the following simple search algorithm. Owing to the appended hash values, the index search complexity is reduced to O (1).

**Search algorithm:**
- Start from head node and go down
- Ignore the appended values and follow reconstructed node.
- Do until level $\log_2|V| - 1$
- Now, finally select the appended value which is the index value

The mode 3 is suitable if the keychain is very long.

### C. Authentication Protocol Suite:

A customer node $C_i$ is authenticated in three different ways. When it joins the system, it goes through a password-based authentication procedure termed the initial authentication protocol. Hereafter, when $C_i$ moves across the network, it is re-authenticated by a certificate-based authentication procedure known as a re-authentication protocol, where a certificate is a time-based ticket issued by $ME_j$ during the initial authentication protocol run.

a) **Initial Authentication Protocol:**

After every $ith$ interval, $P_j$ broadcasts $ME_j$'s public key. A newly joining $C_i$ may receive multiple broadcast messages; however, $C_i$ continues with the first $P_j$, and the protocol proceeds as follows:

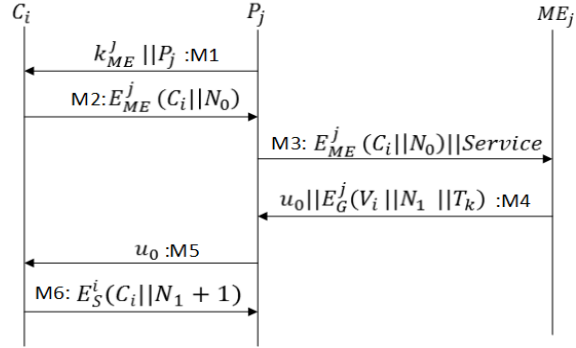

Figure 4. Message exchange for the initial authentication protocol.

M1.      $P_j$ broadcast $ME_j$'s public key.

M2.      In the joining request $C_i$ sends $N_0$ encrypted with $ME_j$'s public key to $P_j$. If $C_i$ is already registered with $ME_j$, the nonce $N_0$ can be replaced with the hash value of the password.

M3.      $P_j$ forwards the request to $ME_j$. $ME_j$ retrieves the profile from the database; if $C_i$ is authorized for the requested services, $ME_j$ retrieves $C_i$'s secret key and sends the message M4.

M4.      $ME_j$ sends M4 to $P_j$ composed of ticket $T_k$, the index value, and $N_1$ (challenge for $C_i$) all encrypted with $K_G^j$ intended for $P_j$ and $u_0 = E_C^i(P_j||N_0 + 1\ ||N_1\ ||T_k\ ||T_R||K_S^i)$ intended for $C_i$. $P_j$ retrieves the customer profile and session key $K_S^i$ from the ticket $T_k$.

M5.      $P_j$ forwards $u_0$ to $C_i$. After challenge verification, $C_i$ accepts $T_k$.
  a. $P_j \rightarrow C_i: u_0 ||Limited\ \therefore$ if the service provider entity cannot fulfill the service requested due to resource limitations, it sends a message 'Limited'. $C_i$ may continue or connect to another service provider entity with the allotted ticket.

M6.      After challenge confirmation, $P_j$ starts providing services; otherwise, $P_j$ halts the service and announces $T_k$ as an invalid ticket.

b) **Re-Authentication Protocol-1:**

When an authenticated $C_i$ moves from $P_k \rightarrow P_j$ such that $\{P_k, P_j\} \in G_i$, the protocol proceeds as follows.

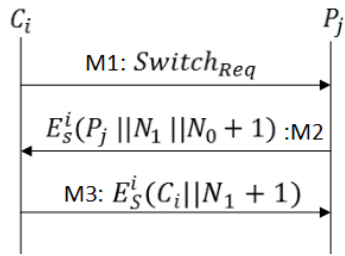

Figure 5. Message exchange for re-authentication protocol-1.

M1.     $C_i$ sends $Switch_{Req} = E_S^i(C_i \,||N_0)||T_k\,||h(ME_i)$ to $P_j$. $P_j$ decrypts the ticket, retrieves the customer profile, and confirms whether $C_i$ is authorized for the further service.

       a. Note that if $P_j$ receives multiple identical $Switch_{Req}$ messages from $C_i$, it indicates the existence of a malicious user.

M2.     $P_j$ sends a challenge response along with a new challenge for $C_i$ encrypted with the C↔P session key.

M3.     After challenge confirmation, $P_j$ starts providing services; otherwise, $P_j$ halts the service and announces $T_k$ as an invalid ticket.

c) **Re-Authentication Protocol-2:**

When an authenticated $C_i$ moves from $P_k \rightarrow P_j$ such that $P_j \in G_k^o$, the protocol proceeds as follows.

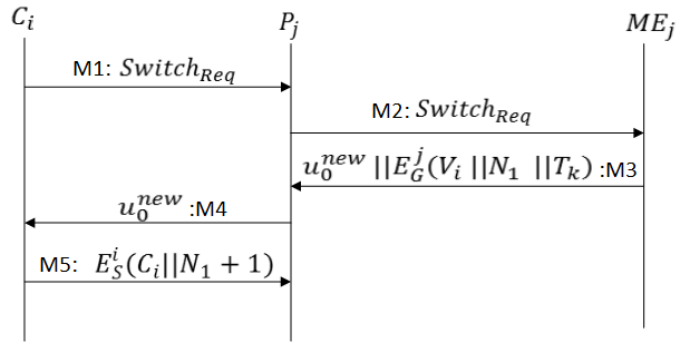

Figure 6. Message exchange for re- authentication protocol-2.

M1.     $C_i$ sends $Switch_{Req} = E_S^i(C_i\,||N_0)||T_k\,||h(ME_i)$ to $P_j$. The $P_j$ decrypts the second half of $T_k$; if the request is not from $C_i$, then $P_j$ discards the request; otherwise, $P_j$ forwards it to $ME_j$.

M2.     $ME_j$ decrypts the ticket, retrieves the customer profile, and confirms whether $C_i$ is authorized for the further service. If $C_i$ is eligible for further services, $ME_j$ generates $K_S^i = h(K_c^i||V_i)$ and proceeds as described below. Otherwise, $ME_j$ ignores the request and $C_i$ may initiate the initial authentication protocol. If $C_i$ sends multiple M1 messages, $P_j$ ignores the messages, and $C_i$ is marked as a malicious user.

M3.     $ME_j$ sends M3 to $P_j$ composed of new ticket $T_k$, the index value $V_i$, and $N_1$ (challenge for $C_i$) encrypted with $K_G^j$ intended for $P_j$ and $u_0^{new} = E_C^i(T_k\,||T_R\,||K_S^i\,||N_1\,||N_0 + 1\,||P_j)$ intended for $C_i$. $P_j$ retrieves the customer profile and session key $K_S^i$ from the new ticket $T_k$.

M4.     $P_j$ forwards $u_0$ to $C_i$. After challenge verification, $C_i$ accepts the $T_k$.

M5.     After challenge confirmation, $P_j$ starts providing services; otherwise, $P_j$ halts the service and announces $T_k$ as an invalid ticket.

If $C_i \notin G_k^o \cup G_k$, the $P_j$ ignores the request, and $C_i$ initiates the initial authentication protocol.

d) **Special Cases:**
**Customer-Customer ($C_i \leftrightarrow C_j$) Mutual Authentication:**

Here, we assume that two customer nodes ($C_j$ and $C_i$) want to communicate directly. To authenticate each other, $C_j$ and $C_i$ exchange messages composed of respective tickets and partial keys which are encrypted with the respective C↔P session keys.

$C_i \rightarrow C_j: E_s^i(C_i||N_0||K_i^p)||T_k$.

$C_j \rightarrow C_i: E_s^j(C_j||N_0+1||N_1||K_j^p)||T_k$.

To decrypt the authentication message, both $C_i$ and $C_j$ forward it to their associated service provider entities. After the first message exchange, there are three possible scenarios concerning the C ↔ P association. The customer-customer mutual authentication protocol proceeds differently for each scenario, as discussed below.

1- If $\{C_i, C_j\}\Delta P_j$, the respective service provider entity ($P_j$) decrypts the messages and retrieves the partial keys for associated customers; $P_j$ sends the partial key along with a challenge response to $C_j$ and $C_i$. After challenge verification, both $C_i$ and $C_j$ generate a C↔C session key $K_{i,j} = H(K_i^p \oplus N_1 || K_j^p \oplus N_0)$ for further communication.

2- If $\{P_i, P_j\} \in G_j$ and $C_i \Delta P_i$ and $C_j \Delta P_j$, the respective service provider entities decrypt the message and retrieve the partial key for associated customers. $P_j$ and $P_i$ send the partial key along with a challenge response to $C_j$ and $C_i$, respectively. After the challenge verification step, both $C_i$ and $C_j$ generate a C↔C session key $K_{i,j} = H(K_i^p \oplus N_1 || K_j^p \oplus N_0)$ for further communication.

3- If $\{P_i, P_j\} \in G_j^0$, $C_i \Delta P_i$, and $C_j \Delta P_j$, the individual service providers forward the message to the respective $ME$s to retrieve the partial keys. After receiving the response from the respective $ME$ s, $P_j$ and $P_i$ send the partial key along with a challenge response to $C_j$ and $C_i$, respectively. After challenge verification, both $C_i$ and $C_j$ generate a C↔C session key $K_{i,j} = H(K_i^p \oplus N_1 || K_j^p \oplus N_0)$ for further communication.

**Delayed or lost response for a joining/switch request:**

As discussed earlier, multiple identical joining/switch requests from the same $C_i$ indicate the existence of an intruder. The inclusion of the previous nonce prevents the situation of misinterpretation between a lost request and a replay attack.

a. $C_i \rightarrow P_j: E_{ME}^j(C_i||N_0||N_0')$.

b. $C_i \rightarrow P_j: E_s^i(C_i||N_0||N_0')||T_k||Switch_{Req}$.

For re-authentication protocol-2, if $C_i$ does not obtain a response for a switch request, it indicates that an intruder has forged $h(ME_i)$, and $P_j$ is unable to proceed. Hence, $P_j$ ignores the request and $C_i$ sends an initial authentication request along with nonce sent in the previous request. The inclusion of the previous nonce prevents the situation of misunderstanding between a lost request and a replay attack..

a. $C_i \rightarrow P_j: E_{ME}^j(C_i||N_0||N_0'||T_k||Alert)$.

## III. Strength against several known attacks

To verify the security of TAP, we introduce an intruder $Z$ into the system, as discussed in earlier work [17]. The intruder is capable of controlling all communication channels (send and receive); it can redirect, spoof, replay or block messages and has initially known information, such as the IDs of all users. In the presence of such an intruder, we explore the strength of TAP against certain known attacks, in this case the replay, parallel session, and binding attacks.

### A. Impersonating $C_i$

Let us consider an intruder $Z(C_i)$ who intercepts and records the messages from $C_i$ and can communicate with $P_j$. During the initial authentication process, if $Z(C_i)$ impersonates $C_i$ without intercepting original messages from $C_i$, at M3 the presence of the intruder is detected, as $ME_j$ receives multiple join requests from the same $C_i$ such that the requests come from multiple or single instances of $P_j$ with multiple requests per $P_j$, thus indicating the presence of an intruder. In the given scenario, $ME_j$ sends an alert message appended with M4; additionally, each M4 includes a different $N_1$ challenge nonce for each request forwarding $P_j$ ( $ME_j \rightarrow P_j : u_0 \; ||E_G^j(V_i \; ||N_1 \; ||T_k)||Alert$ ). Similarly, even if $Z(C_i)$ impersonates and successfully intercepts all messages from $C_i$, it still fails to send M6 without knowing the C↔P session key and the private key of $C_i$. In re-authentication protocol-1 and protocol-2, if $Z(C_i)$ impersonates and successfully intercepts all the messages from $C_i$, without knowing the C↔P session key, $Z(C_i)$ fails to send a valid M3 in protocol-1 and an authentic M5 in protocol-2. The failure of M3/M5 indicates the presence of a malicious user. Likewise, in mutual authentication protocol for customers, an intruder cannot acquire partial keys without knowing the C ↔ P session keys.

### B. Impersonating $P_i$

Let us consider an intruder $Z(P_i)$ who can intercept and record messages from $P_j$ and can communicate with $C_i$. If $Z(P_j)$ impersonates $P_j$ and successfully intercepts all messages from $P_j$, it brings a minor delay, and $C_i$ initiates another authentication process. Moreover, $Z(P_j)$ is unable to obtain $T_k$ without knowing the private key of $C_i$ and cannot obtain $K_G^j$ without knowing the private key of $P_j$. Similarly, in re-authentication protocol-1 and protocol-2, if $Z(P_j)$ impersonates and successfully intercepts all messages from $P_j$, $Z(P_j)$ is still unable to obtain $T_k$ without knowing the C↔P session key.

### C. Replay or Multiplicity Attack

To investigate the strength of TAP against a replay or multiplicity attack, as introduce the intruder $Z(C_i)$ , as discussed above, and launch a replay attack against initial and re-authentication protocols.

a) **Replay attack on the Initial Authentication Protocol**

In the $\{Z(C_i), C_i\}\Delta ME_j$ case, the intruder $Z(C_i)$ can replay a few messages in the initial authentication protocol run; however, it fails to complete the protocol run, and the presence of the intruder is detected after few messages are exchanged.

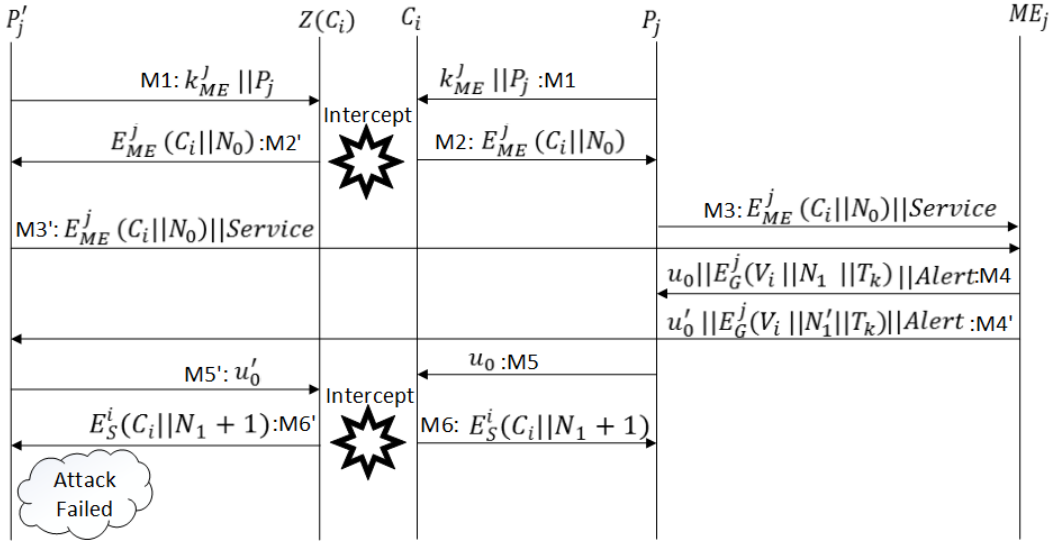

Figure 7. An example of replay attack on TAP initial authentication protocol.

- $M2'$: Intruder $Z(C_i)$ intercepts M2 and replays the message towards $P_j'$.
- $M3'$: $ME_j$ receives similar multiple join requests from $P_j$ and $P_j'$, which is a conceivable indication of a malicious user.
- $ME_j$ will send $M4$ and $M4'$ to $P_j$ and $P_j'$, respectively. Both messages consist of an alert signal and two different challenges.
- $M6'$: The intruder fails to complete the attack, as the expected reply is $E_S^i(C_i||N_1'+1)$, and $P_j'$ notifies $ME_j$ of the existence of an intruder.

b) **Replay attack on Re-Authentication Protocol-1**

In the case of $\{P_j', P_j\}\Delta ME_j$, the intruder $Z(C_i)$ can replay a few messages in re-authentication protocol-1; however, it fails to complete the protocol run and the presence of the intruder is thus detected after a few messages are exchanged.

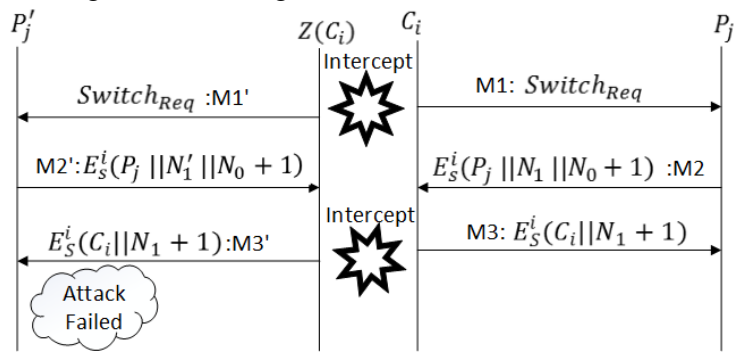

Figure 8. An example of replay attack on TAP re-authentication protocol-1.

- $M1'$: The intruder replays the message to deceive $P_j'$ with a message seemingly sent by $C_i$ and encrypted with $K_s^i$.
- $M3'$: The intruder fails to complete the attack, as the expected reply is $E_S^i(C_i||N_1'+1)$, and $P_j'$ notifies $ME_j$ of the existence of an intruder.

### c) Replay attack in Re-Authentication Protocol-2

In the case of $\{Z(C_i), C_i\}\Delta ME_j$, the attack follows a message sequence similar to that of the initial authentication case and provides the same conclusion; at M3, $ME_j$ detects the threat, and at M5 the corresponding $P_j$ identifies the intruder. However, if $\{C_i\}\Delta ME_j$ and $Z(C_i)\Delta ME_k$ such that $\{ME_j, ME_k\} \in G_j$, meaning that the intruder intercepts the messages and uses it to become authenticated by another $ME_j$ by exploiting the fact that the $ME_js$ do not interact throughout the procedure. This attack is carried out as follows:

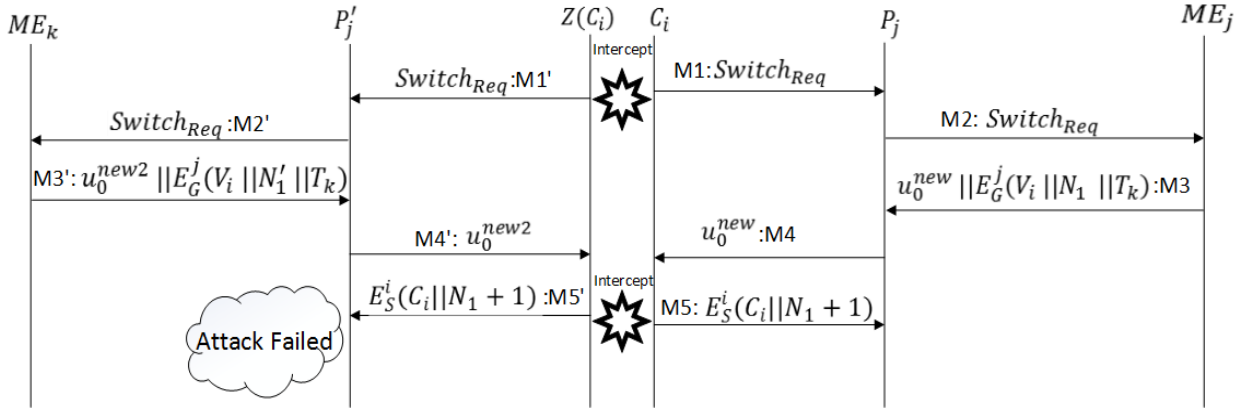

Figure 9. An example of replay attack on TAP re-authentication protocol-2.

- $M1'$: The intruder replays the message to deceive $P_j'$ with a message seemingly sent by $C_i$ and encrypted with $K_s^i$.
- $M3/M3'$: $ME_j$ or $ME_k$ cannot detect an intruder.
- $M5'$: The intruder fails to complete the attack, as the expected reply is $E_S^i(C_i||N_1'+1)$, and $P_j'$ notifies $ME_j$ of the existence of an intruder

### D. TAP failure under given conditions

Let's Next, we consider a certain exceptional condition which enables an intruder to launch a successful attack.

C1. $Z(C_i)$ is in $P_j$ coverage area.
C2. $C_i$ is not in the range of any legitimate $P_j$.
C3. $Z(P_j)$ can communicate with $C_i$ pretending to be a legitimate service provider.
C4. $Z(C_i)$ and $Z(P_j)$ can communicate directly with negligible time delay.
C5. All four conditions should hold together.

Under the given conditions, the intruder can act as a 'man in middle', replay messages, and run a parallel session of the protocol with 'binding attack' abilities. The intruder replays the message which it receives from its partner, who runs a parallel session with a legitimate counterpart. However, even under the given conditions, the intruder cannot re-authenticate. The attack on the initial authentication will proceed as follows:

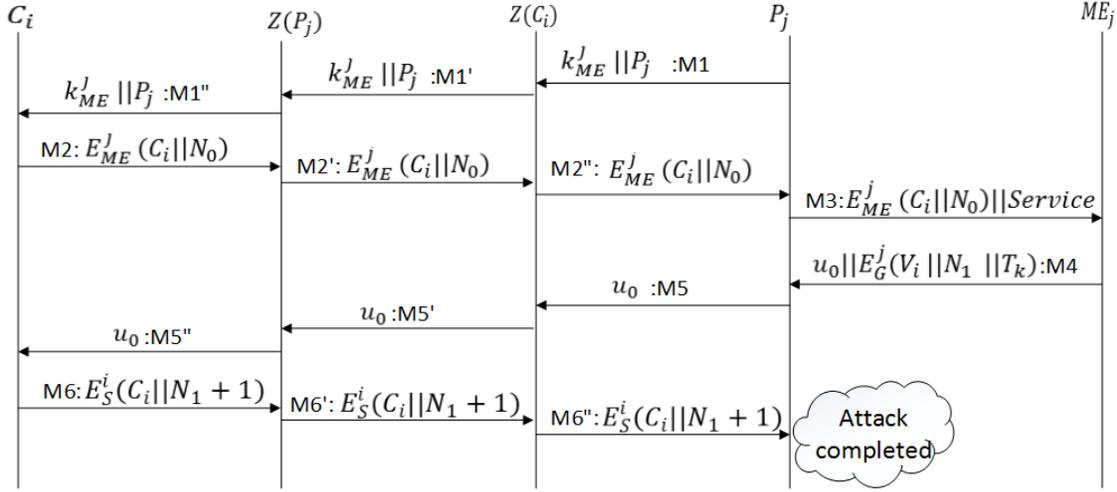

Figure 10. An example of a successful attack on TAP under given conditions.

### E. Remarks on TAP security

In TAP, if a legitimate counter entity is an active participant in the system, the impersonating $C_i$ or $P_j$ is detectable and identifiable; this implies that replay, binding, and parallel session attacks are not successful attacks against TAP. However, under highly exceptional conditions, a man in the middle, who can replay the messages and run a parallel session of the protocol with the capability of a binding attack, can launch a successful attack. In a system where these conditions are likely to occur, it is recommended to take certain measures to authenticate the identity of the broadcasting service provider entity ($P_j$). To prevent DDoS attacks, the group join and initial authentication procedures include a puzzle in M1 and proceed if the requesting entity provides a valid solution [18,19].

### VII. Formal analysis of TAP using CSP and RANK function analysis

Communicating Sequential Processes (CSP) refer a form of algebra which describes and analyzes a system which consists of communicating processes [20]. TAP authentication procedures can be defined as events of CSP processes, and through the rank function analysis method [21-23], we can verify the security of TAP.

### CSP Notations:
- $A \rightarrow P$: Process $P$ performs an event $A$ on its interface. After performing event $A$ process $P$ may or may not change its state.
- $P \square Q$: $P$ Choice $Q$ is a choice operator, it provides a process who behaves as $P$ or $Q$.
- $P|R|Q$: $P$ and $Q$ run in parallel and synchronized on event $P$.

- $P|R|STOP$: Restrict process $P$ on event set $R$.
- $P|||Q$: $P$ and $Q$ are the interleaving processes, and run in parallel without event synchronization. It is a special case of the parallel process, where an event set of synchronization is an empty set.
  - $P|||Q = P|<>|Q$.
- $S \vdash m$: Set of messages $S$ can generate message $m$.
- $traces(P)$: All possible event traces of process $P$.
  - $tr \in traces(P)$: A trace sequence $tr$ belongs to $traces(P)$ if $P$ performs events of $tr$ in the same sequence.
- $tr \Downarrow C$: The set of messages in trace sequence $tr$ collected at channel $C$.
- $tr \upharpoonright A$: Maximal subsequence of trace sequence $tr$ who's elements are taken from event set $A$.
- $P\ sat\ S \Leftrightarrow tr \in traces(P) \circ S$.
  - If the trace $tr$ is one of the traces of process $P$ such that trace $tr$ predicated by the event/message $S$, implies that process $P$ satisfies the event/message set $S$. Such statement is trace specification (TS).
- $CP_{id}$: Set of id's of all $C_i$ and $P_j$.
- $n \approx m$: $n$ is sufficient information to trust that $m$ is correct/true information.

### A. Modeling the TAP network in CSP

We consider the system $NET$, which is defined by the legitimate user processes of TAP in conjunction with the intruder process $Z$.

$$NET = (User_C^i(N_0)\ |||User_P^j\ |||User_{ME}^j)\ |[send, receive]|User_Z(S)$$

Where, $User_C^i(N_0)$, $User_P^j$ and $User_{ME}^j$ represent the legitimate interleaving processes of $C_i$, $P_j$ and $ME_j$, respectively, which are running in parallel and which are synchronized to intruder $Z$ at the $[send, recieve]$ event set. This gives the intruder the capabilities discussed below.

#### a) Modeling Intruder $Z$ in CSP

Let us define a CSP model for intruder process $Z$ as discussed in an earlier study [17] with certain extra capabilities. The intruder is capable of controlling all communication channels (send and receive) and can redirect, spoof, replay or block messages. It also has the initially known information $INIT$, e.g., the IDs of all users. When a legitimate user sends a message to another legitimate user, $Z$ can intercept and record each message; these intercepted messages are additional information with regard to the current information $(S)$ of $Z$ i-e $S \cup \{m\}$.

$$User_Z(S) = \quad send.A.B.m \rightarrow Z(S \cup \{m\})$$
$$\square_{S \vdash m}\ receive.A.B.m \rightarrow Z(S)$$

Let us also suppose the existence of legitimate users $A$ and $B$. Intruder $Z$ can deceive $A$ by sending a message $m$, intercepted or generated from current information $S$ ($S \vdash m$), and pretending to be user $B$.

### b) Modeling TAP Authentication Procedures in CSP

The activities of $C_i$, $P_j$ and $ME_j$ in TAP can be defined as CSP processes. In addition to the activation of TAP authentication, the CSP processes also generate two signals, $conf$ and $Auth$:

- $conf.A.B.n \approx m$ : Based upon information $n$, participant A is confident that shared information $m$ is trusted information between A and B.
- $Auth.B.A.m$: B authenticates A based upon trusted information $m$ and B agrees that A was previously running the protocol and performed the corresponding signal $conf$.

The CSP processes for the TAP initial authentication protocol:

$$User_C^i = \square_P^j \; receive.C_i.P_j.K_{ME}^j \rightarrow$$
$$send.C_i.P_j.E_{ME}^j(C_i||N_0) \rightarrow$$
$$receive.C_i.P_j.u_0 \rightarrow$$
$$conf.C_i.P_j.N_0 \approx N_1 \rightarrow$$
$$send.C_i.P_j.E_s^i(C_i||N_1+1) \rightarrow STOP$$

$$User_P^j = send.P_j.C_i.K_{ME}^j \rightarrow$$
$$receive.P_j.C_i.E_{ME}^j(C_i||N_0) \rightarrow$$
$$send.P_j.ME_j.E_{ME}^j(C_i||N_0) \rightarrow$$
$$receive.ME_j.P_j.u_0 \rightarrow$$
$$send.P_j.C_i.u_0 \rightarrow$$
$$receive.P_j.C_i.E_s^i(C_i||N_1+1) \rightarrow$$
$$Auth.P_j.C_i.N_1 \rightarrow STOP$$

The CSP processes for TAP re-authentication protocol-1:

$$User_C^i = \square_P^j \; send.C_i.P_j.E_s^i(C_i||N_0)||T_k \rightarrow$$
$$receive.C_i.P_j.E_s^i(P_j||N_1||N_0+1) \rightarrow$$
$$conf.C_i.P_j.N_0 \approx N_1 \rightarrow$$
$$send.C_i.P_j.E_s^i(C_i||N_1+1) \rightarrow STOP$$

$$User_P^j = receive.P_j.C_i.E_s^i(C_i||N_0)||T_k \rightarrow$$
$$send.P_j.C_i.E_s^i(P_j||N_1||N_0+1) \rightarrow$$
$$receive.P_j.C_i.E_s^i(C_i||N_1+1) \rightarrow$$
$$Auth.P_j.C_i.N_1 \rightarrow STOP$$

The CSP processes for TAP re-authentication protocol-2:

$$User_C^i = \square_P^j \; send.C_i.P_j.E_s^i(C_i||N_0)||T_k||h(ME_i) \rightarrow$$
$$receive.C_i.P_j.u_0 \rightarrow$$
$$conf.C_i.P_j.N_0 \approx N_1 \rightarrow$$
$$send.C_i.P_j.E_s^i(C_i||N_1+1) \rightarrow STOP$$

$$User_P^j = receive.P_j.C_i.E_s^i(C_i||N_0)||T_k||h(ME_i) \rightarrow$$
$$send.P_j.ME_j.E_s^i(C_i||N_0)||T_k||h(ME_i) \rightarrow$$
$$receive.ME_j.P_j.u_0 \rightarrow$$
$$send.P_j.C_i.u_0 \rightarrow$$
$$receive.P_j.C_i.E_s^i(C_i||N_1+1) \rightarrow$$
$$Auth.P_j.C_i.N_1 \rightarrow STOP$$

Where the activities of $ME_j$ are common for all of the TAP protocols.

$$User_{ME}^j = receive.ME_j.P_j.(C_i||N_0) \rightarrow$$
$$send.ME_j.P_j.u_0||E_G^j(V_i||N_1||T_k) \rightarrow STOP$$

### B. Rank Function Analysis

To prove the overall security of TAP, we use a proof strategy which verifies that all authentication protocols satisfy the rank theorem, which implies that TAP is a secure protocol. The entire process is discussed in detail in subsequent sub-sections.

### a) Proof strategy

According to the protocol semantics, the signal $conf.A.b.m$ must follow a signal $Auth.A.B.m$, which implies the following authentication property (AP1):

$$NET \; Sat \; conf.C_i.P_j.(N_0 \approx N_1) \; precedes \; Auth.P_j.C_iN_1$$

To prove that $NET$ should meet AP1, Schneider [22] specifies a simple strategy: for $NET$ to satisfy AP1, $NET$ must establish that $Auth.P_j.C_iN_1$ cannot be generated in $NET$ if an occurrence of $conf.C_i.P_j.N_0 \approx N_1$ is prevented. This proof strategy implies the following trace specification (TS1):

$$NET \,|conf.C_i.P_j.N_1|\, STOP \; Sat \; tr{\upharpoonright}Auth.P_j.C_iN_1 = <\;>$$

**b) Rank Function**

Rank function ($\rho : M \to Num$) maps the messages to a number, where $M$ is the set of all messages and signal generated messages ($S \vdash m$) appearing in the protocol run. $\rho(m) > 0$ if the disclosure of $m$ is safe (i.e., if $NET$ maintains a secure state) and $\rho(m) \leq 0$ if the disclosure of $m$ is unsafe (i.e., if $NET$ enters a compromising state). For a process $P$ to maintain a positive $\rho$, it should not transmit $\rho(m) \leq 0$ until and unless $P$ has already received $(m) \leq 0$. Such a process maintains the following trace specification (TS2).

$$P \; maintains \; \rho \iff \forall \; tr \in trace(P) \circ \rho(tr \Downarrow receive) > 0 \Rightarrow \rho(tr \Downarrow send) > 0$$

If a process maintains TS2, it never introduces $\rho(m) \leq 0$; hence, a protocol is proved to be secure if all processes maintain TS2.

**c) Rank Theorem and Sufficient Condition**

**Rank Theorem**:

If, for event sets $R$ and $T$, $\rho$ satisfying

P1) $\forall m \in INIT \circ \rho(m) > 0$

P2) $\forall S \subseteq M, \; m \in M \circ (\forall m' \in S \circ \rho(m') > 0) \wedge S \vdash m \Rightarrow \rho(m) > 0$

P3) $\forall t \in T \rho(t) \leq 0$

P4) $(User_i|R|stop \; sat \; maintain \; \rho)$ for each user $i$

then, $NET \; sat \; R \; Precedes \; T$.

The proof of theorem is presented in earlier work [20]. The four properties of the rank function prevent an exchange of non-positive messages in the system $NET|R|STOP$, which is synchronized with $STOP$ upon event set $R$. In TAP, $R = conf.A.B.m$ and $T = Auth.A.B.m$, which implies that the occurrence of signal $conf$ will stop the process $NET$.

Earlier, we discussed with regard to the proof strategy that TAP is secure if it satisfies TS1. If the rank function of TAP holds all properties of the rank theorem, this implies that all of the processes maintain TS2, and $NET$ satisfies TS1. This condition is sufficient to prove that TAP is a secure authentication protocol.

**d) Rank Analysis for TAP initial Authentication**

Here, we define the rank function for the initial authentication. As discussed earlier, if an intruder impersonates $C_i$ or $P_j$, the system $NET$ maintains a secure state; hence, $\rho(CP_{id}) > 0$. All nonce instances are non-positive such that $\rho(N) \leq 0$; hence, the nonce must be sent out encrypted with a secure key ($K$ represents the set of all secure keys); i.e., $\rho(E_k(N)) > 0$. As described in TS1, $NET$ is restricted upon $conf.A.B.m$, and any message or signal after the $conf$ signal is marked as non-positive.

$$\rho(K) = \begin{cases} 1 & if\ K = K_{ME}^j \\ 0 & otherwise \end{cases}$$

$$\rho(m) = \begin{cases} 0 & if\ m = receive.P_j.C_i.E_s^i(C\_i||N_1 + 1)\ Or \\ & m = send.C_i.P_j.E_s^i(C_i||N_1 + 1) \\ 1 & otherwise \end{cases}$$

$$\rho(CP_{id}) > 0$$
$$\rho(E_k(N)) > 0$$
$$\rho(Athu) \leq 0$$
$$\rho(N) \leq 0$$

At this stage, we can verify the security of the TAP initial authentication using the rank theorem properties.

**P1)** $\forall m \in INIT \circ \rho(m) > 0$: The fundamental knowledge of the intruder $INIT$ includes $CP_{id}$ and $K_{ME}^j$, which have positive rank values, hence satisfying P1.

**P2)** $\forall S \subseteq M, m \in M \circ (\forall m' \in S \circ \rho(m') > 0) \wedge S \vdash m \Rightarrow \rho(m) > 0$ : This property verifies whether a set of positive rank messages can generate a non-positive rank message. All non-positive rank messages are encrypted with the non-positive encryption key $K_s^i$. The intruder cannot acquire $K_s^i$ without knowing the non-positive key $K_c^i$, the non-positive nonce $N_0$, and the non-positive nonce $N_1$; hence, this condition satisfies P2.

**P3)** $\forall t \in T \rho(t) \leq 0$: The event set $T$ is the $Auth.A.B.m$ signal, and under the given restriction we have $\rho(Athu.A.B.m) \leq 0$, hence satisfying P3.

**P4)** $(User_i|R|stop\ sat\ maintain\ \rho)$ for each user

This property states that all users $i \in CP_{id}$ should maintain a positive state when restricted with regard to event set $R = conf.A.B.m$. We confirm if the processes $User_C^i$ and $User_P^j$ satisfy P4

$$User_C^i|Conf.C.P.N_1|STOP = \square_P^j\ receive.C_i.P_j.k_{ME}^j \rightarrow$$
$$send.C_i.P_j.E_{ME}^j(C_i||N_0) \rightarrow$$
$$receive.C_i.P_j.u_0 \rightarrow$$
$$if\ P = P_j \wedge N = N_0 \approx N_1\ STOP$$
$$else\ if\ P = P_j \wedge N \neq N_0$$
$$Initiate\ with\ resend\ mode$$

Hence, the protocol maintains a positive $\rho$ under the given restriction, and it satisfies P4.

It is concluded that the TAP initial authentication protocol is secure, and we further check re-authentication protocol-1 and protocol-2.

### e) Rank Analysis for TAP Re-Authentication

Rank function for TAP re-authentication protocols 1 and 2 are similar to the rank function for initial authentication (because all the conventions of initial authentication hold true for re-authentication protocol 1 and 2), except ticket, which is encrypted with non-positive keys, hence we have $\rho(T_k) > 0$. From CSP of TAP re-authentication, we notice that message and signal pattern after $conf.A.b.m$ is similar to the initial authentication protocol. Hence, TAP satisfies the rank properties for re-authentication protocol 1 and 2 as well. It concludes that TAP is a secure authentication protocol.

## VIII. Security and Performance Comparison

### A. Security Comparison

In this section, we compare TAP with existing authentication schemes [24-27]. The TAP protocol is not limited to a particular network type or application scenario; thus, we compare TAP with a sensor network [24-25], LTE [26] and client-server applications [27]. In the previous sections, we examined the strength of TAP with a rigorous analysis. For further confirmation of the strength of the TAP protocol, we implemented TAP and several well-known previously proposed schemes [24-27] in an automated security protocol analysis tool, Scyther [28]. The Scyther tool verifies protocol claims against possible attacks in the presence of an intruder, as discussed in Section V-B. The claims are events which describe the aim and security properties of the authentication protocol, as defined below [1-2].

**Aliveness:** This claim infers that at the end of a protocol run, the participants are guaranteed that all participants were running the protocol. **Weak Agreement:** This claim presumes that at the end of the protocol run, protocol initiator is confident that the protocol responder has been running the protocol, though superficially. **Non-injective Agreement:** This claim infers that at the end of a protocol run, the protocol initiator is confident that the protocol responder has been running the protocol according to a defined role and participants are agreed upon a data set shared during the protocol run. **Non-injective Synchronization:** This claim infers that at the end of a protocol run, all participants are confident that all other participants exactly followed their roles in the protocol and exchanged messages in the intended order.

TABLE I
SCYTHER PARAMETER SETTINGS

| Parameter | Type |
|---|---|
| Number of Runs | 1~100 |
| Matching Type | Find all Type Flaws |
| Search pruning | Find All Attacks |
| Number of pattern per claim | 100 |

TABLE II
AUTHENTICATION PROPERTIES COMPARISON

| Claims | B. Vaiyda et. al [24] | | | | I. Chang et. al [25] | | | | Li Xiehua [26] | | | | | | H. Lin et. al [27] | | | | TAP | | | TAP | | |
|---|---|---|---|---|---|---|---|---|---|---|---|---|---|---|---|---|---|---|---|---|---|---|---|---|
| | Not Restricted | | Restricted | | Not Restricted | | Restricted | | Not Restricted | | | Restricted | | | Not Restricted | | Restricted | | Not Restricted | | | Restricted | | |
| | U | GW | U | GW | U | GW | U | GW | UE | MME | HSS | UE | MME | HSS | U | S | U | S | C | P | ME | C | P | ME |
| Aliveness | N | Y | N | Y | Y | Y | Y | Y | N | N | N | N | Y | N | N | Y | N | Y | Y | Y | Y | Y | Y | Y |
| Weak Agreement | N | Y | N | Y | Y | Y | Y | Y | N | N | N | N | Y | N | N | Y | N | Y | Y | Y | Y | Y | Y | Y |
| Non-injective Agreement | N | Y | N | Y | Y | Y | Y | Y | N | N | N | N | Y | N | N | Y | N | Y | N | Y | Y | Y | Y | Y |
| Non-injective Agreement | N | Y | N | Y | Y | Y | Y | Y | N | N | N | N | Y | N | N | Y | N | Y | N | N | N | Y | Y | Y |

In Scyther the protocols are modeled as an exchange of messages among the participants performing specific 'roles'; for instance, the customer node performs the role of the initiator, the service provider performs the role of the responder, and the ME performs the role of a server. We implemented and tested TAP and the proposed methods of Vaiyda et al. [24], Chang et al. [25], Xiehua [26], and Lin et al. [27] through the claims mentioned above with the parameter settings given in Table I.

The protocols are tested under 'Restricted' and 'Not Restricted' conditions. Under the 'Restricted' conditions, honest participants using the protocol are restricted and can thus run only one instance of the protocol. These results are shown in Table II. It is clear that our protocol qualifies all of the protocol claims, and no attacks were noted under the restricted condition. Conversely, it fails to fulfill a few claims when participants are permitted to run multiple instances. However, our protocol outperforms those in the earlier works [24-27], and it is secure in a large number of systems and scenarios. In contrast, the earlier methods [24-27] are susceptible to several attacks and fail to fulfill the majority of authentication claims.

### B. Performance Comparison

The results of the performance comparison are presented in Table III, where we compare the efficiency of the TAP authentication protocol suite in terms of the computational cost, message complexity and time synchronization requirements against the authentication schemes discussed above [24-27]. The computational cost is estimated to be the sum of the overall number of modular exponentiations (e) and the hash (h) and XOR (x) operations. To compute the computational cost of one of the earlier methods [26], we assumed that the cost of functions $f3$, $f4$, and $s10$ in the SE-EPS vector generation algorithm were identical to one hash operation. Regarding the computational cost, the TAP protocol suite greatly outperforms all of the schemes. Referring to the modular exponentiation, the approach presented by Lin et al. [27] is the most expensive scheme, followed by those of Chang et al. [25], Xiehua [26] and Vaiyda et

TABLE III
PERFORMANCE COMPARISON OF TAP WITH PREVIOUS WORK

| Scheme↓ | Comp. Complexity | Message Complexity | Time Syn. |
|---|---|---|---|
| B. Vaiyda et. al [24] | 8H+4X | (6+k)U | Y |
| I. Chang et. al [25] | 25H+1X | (2+5)U | Y |
| Li Xiehua [26] | 12H+2X | 8U | N |
| H. Lin et. al [27] | 13H+12X+2E | (2+3)U | Y |
| TAP (IA) | 3H | 5U | N |
| TAP (RA-1) | 1H | 3U | N |
| TAP (RA-2) | 4H | 4U | N |

E=Modular exponentiation, $H$ = hash operation, X=XOR operation, U = unicast message al. [24]. For the sake of simplicity, we ignore the computational cost of the XOR operation. Figure 11 shows the computational cost of TAP compared to the approaches of Vaiyda et al. [24], Chang et al. [25] and Xiehua [26] for a mobile customer node moving across the network and experiencing the authentication process. Figure 12 shows the message complexity of TAP compared to these earlier schemes [24-27] for a mobile customer node moving across the network and experiencing the authentication process multiple times. The message complexity is presented for the method of Vaiyda et al. [24], calculated with the assumption that k=3, meaning that at the time of joining there are five potential nodes which can process the login request sent by a user. In the best-case scenario, when a customer experiences re-authentication protocol 1 the, TAP message complexity is the lowest. However, the message complexity of the approach by Lin et al. [27] is slightly better than the message complexity of TAP in the worst-case scenario. Moreover, unlike the methods of Vaiyda et al. [24], Chang et al. [25] and Lin et al. [27], TAP and the approach by Xiehua [26] do not require time synchronization among the participating entities.

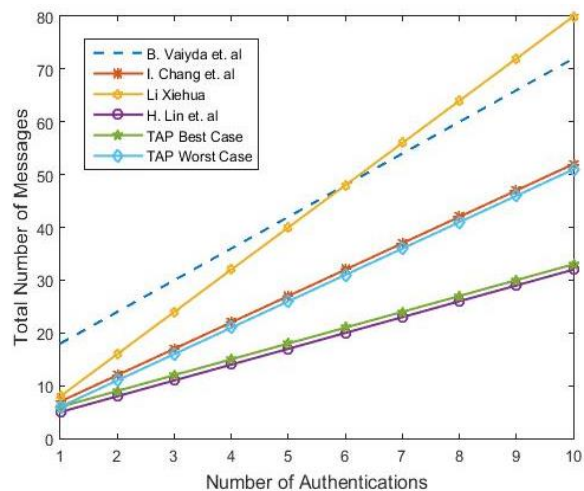

Figure 12. Message Complexity comparison of TAP with previous works.

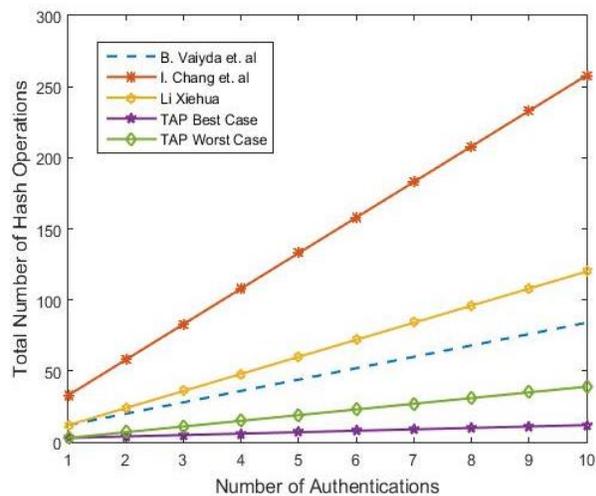

Figure 11. Computational cost comparison of TAP with previous works.

## IX. Conclusion

In this paper, we proposed a novel key distribution and authentication protocol (TAP) for dynamic and mobile network applications. TAP enhances the level of protocol security with the assistance of time-based encryption keys and scales down the authentication complexity by issuing an authentication ticket. A security analysis conducted here shows that TAP is secure against known attacks. A formal analysis using CSP and rank function analysis further confirms the strength of the TAP protocol. We also compared the security and performance of TAP with a sensor network [24-25], LTE [26] and with the Client Server Application approach [27]. The final results show that TAP is secure and desirable for an immense range of network applications.

**ACKNOWLEDGEMENT**
This work was supported by Institute for Information &communications Technology Promotion (IITP) grant funded by the Korea government (MSIP). (No. R0166-16-1017, Standardization of Wireless Power Transfer Technology and Service)]